\documentclass[aps,prb,twocolumn]{revtex4-2} 

\usepackage{array}
\usepackage{amsthm}
\usepackage{graphicx} 
\usepackage{siunitx}
\usepackage[T1]{fontenc}
\DeclareSIUnit[number-unit-product = {}]{\inch}{\textquotedbl}
\usepackage{float}
\usepackage{xcolor}
\usepackage{booktabs}
\usepackage{amsmath}
\usepackage{amssymb}
\usepackage[super]{nth}
\usepackage{hyperref}
\usepackage[capitalize]{cleveref}
\hypersetup{
    colorlinks,
    citecolor=blue,    filecolor=blue,
    linkcolor=blue,    urlcolor=blue
}

\begin{document}

\title[]{All-optical Compton scattering at shallow interaction angles}

\author{A. Döpp}
\address{Ludwig-Maximilians-Universit\"at M\"unchen, Am Coulombwall 1, 85748 Garching, Germany}
\author{K. Ta~Phuoc}
\author{I.~A.~Andriyash}
\address{Laboratoire d’Optique Appliqu\'ee, ENSTA Paris, CNRS, Ecole Polytechnique, Institut Polytechnique de Paris, 828 Bd des Mar\'echaux, 91762 Palaiseau, France}

\begin{abstract}

All-optical Compton sources combine laser wakefield accelerators and intense scattering pulses to generate ultrashort bursts of backscattered radiation. The scattering pulse plays the role of a small-period undulator ($\sim\SI{1}{\micro\meter}$) in which relativistic electrons oscillate and emit x-ray radiation. To date, most of the working laser-plasma accelerators operate preferably at energies of a few hundreds of MeV and the Compton sources developed so far produce radiation in the range from hundreds of keV to a few MeV. However, for such applications as medical imaging and tomography the relevant energy range is $10-100$ keV. In this article, we discuss different scattering geometries for the generation of X-rays in this range. Through numerical simulations, we study the influence of electron beam parameters on the backscattered photons. We find that the spectral bandwidth remains constant for beams of the same emittance regardless of the scattering geometry. 
A shallow interaction angle of 30 degrees or less seems particularly promising for imaging applications given parameters of existing laser-plasma accelerators. Finally, we discuss the influence of the radiation properties for potential applications in medical imaging and non-destructive testing.

\end{abstract}
\maketitle

\subsection{Introduction}

All-optical Compton sources are novel compact X-ray sources, which combine both laser-plasma accelerators (LPAs) and optical undulators for radiation generation, see \citet{Corde.2013} and \citet{Albert.2016} for in-depth reviews. In such devices, electrons are injected and accelerated in the wake of an intense femtosecond laser pulse \cite{Esarey.2009}. The electric fields in this wake reach a few hundreds GeV/meter and during the interaction electrons are accelerated to relativistic energies ($>100$ MeV) on a millimeter scale. Laser-wakefield accelerators with the optical \cite{Faure.2006, Wenz.2019} or density-transition injection \cite{Buck.2013, Goetzfried.2020} have shown great potential to produce stable electron beams with small energy spread. Electrons beams with 1 percent energy spread and 1 mrad divergence can now be reliably produced in the 100-300 MeV range and can be used for the production of x-ray radiation via Compton Backscattering in an all-optical setup \cite{Khrennikov.2015, Powers.2014, Phuoc.2012}.

The theory of Compton backscattering is fundamentally related to synchrotron radiation emission \cite{Corde.2013}. The up-shifted energy of backscattered photons $\hbar\omega$ can be described using an adapted form of the well-known undulator equation \cite{Wiedemann.2007}
\begin{equation}\label{eq1}
 \frac{\hbar\omega}{\hbar\omega_0} = \frac{(2\gamma^2(1-\beta \cos\varphi))}{1+a_0^2/2+\theta^2\gamma^2}.
\end{equation}
Here, $\hbar\omega_0$ is the initial energy of the backscattered photon (taking the position of energy associated with the undulator wavelength), $a_0$ is the normalized peak potential of the scattering laser pulse (analogous to the undulator or wiggler parameter $K$ in conventional light sources) and $\theta$ is the angle of observation close to the axis. Next, $\gamma$ and $\beta=v_e/c_0$ denote the Lorentz factor and velocity of the electrons normalized to the speed of light in vacuum, respectively. Furthermore, optical undulators offer an additional degree of freedom not found in conventional undulators -- the angle $\varphi$ between the electron beam and the undulating structure. It appears in the numerator of \cref{eq1} as the relativistic doppler-shift $(1-\beta \cos\varphi)$. For intense scattering pulses ($a_0\gtrsim 1$) the probability for multi-photon scattering increases and a single electron may absorb $n$ laser photons before emission of a single photon of the energy $n\hbar\omega$ \cite{Yan.2017}.

In the case of backscattering with a head-on collision ($\varphi = \SI{180}{\degree}$), weak scattering potential ($a_0\ll1$) and on-axis observation $\theta=\SI{0}{\degree}$, \cref{eq1} simplifies to
\begin{equation}\label{eq2}
\frac{\hbar\omega}{\hbar\omega_0} = 4\gamma^2.
\end{equation}
Most laser wakefield accelerators rely on titanium:sapphire (Ti:Sa) lasers \cite{Danson.2019} with typical pulse durations of the order of $\SI{30}{\fs}$. To match the pulse duration with the plasma wavelength \cite{Malka.2002,Ding.2020}, the accelerators will typically operate at plasma densities of $n_e\sim10^{19}\si{\per\cubic\cm}$ or lower, resulting in dephasing limited beam energies above 100 MeV \cite{Doepp.2016}. However, following \cref{eq2}, a 100 MeV electron beam will generate backscattered photons at about 250 keV. This is too high for many applications of X-rays, such as  clinical imaging and tomography, which typically require photon energies in the range of $50-100$ keV \cite{Goetzfried.2018}. Reaching lower electron beam energies with Ti:Sa lasers is in principle possible when accepting an inefficient truncation of the acceleration process, but requires sub-millimeter nozzles that are difficult to manufacture. Alternatively, one can shift the regime of operation to higher density by using more complex laser technologies delivering even shorter pulses such as light-field synthesizers \cite{Schmid.2009} or post-compressors using self-phase modulation for spectral broadening \cite{Guenot.2017,Salehi.2021}. 

As we will discuss in this paper, one can take advantage of the scattering geometry to generate the desired sub-100-keV photon beams while still operating at the ''sweet spot`` of LWFA of 100 MeV or more. We can include the scattering angle $\varphi$ again in \cref{eq2} and, for highly relativistic electron beams ($\beta\approx 1$), arrive at the simple expression
\begin{equation}\label{eq:PhotonEnergy}
 \frac{\hbar\omega}{\hbar\omega_0}=4\gamma^2\sin^2\frac{\varphi}{2}.
\end{equation}

\begin{figure}[tb]
\includegraphics[trim=5 0 -50 0, clip,width=.9\linewidth]{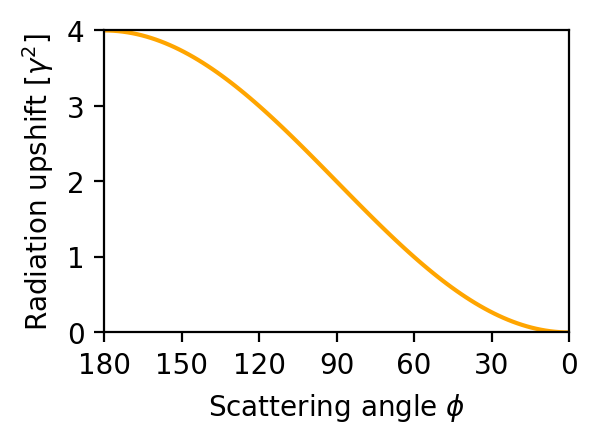}
\includegraphics[width=.98\linewidth]{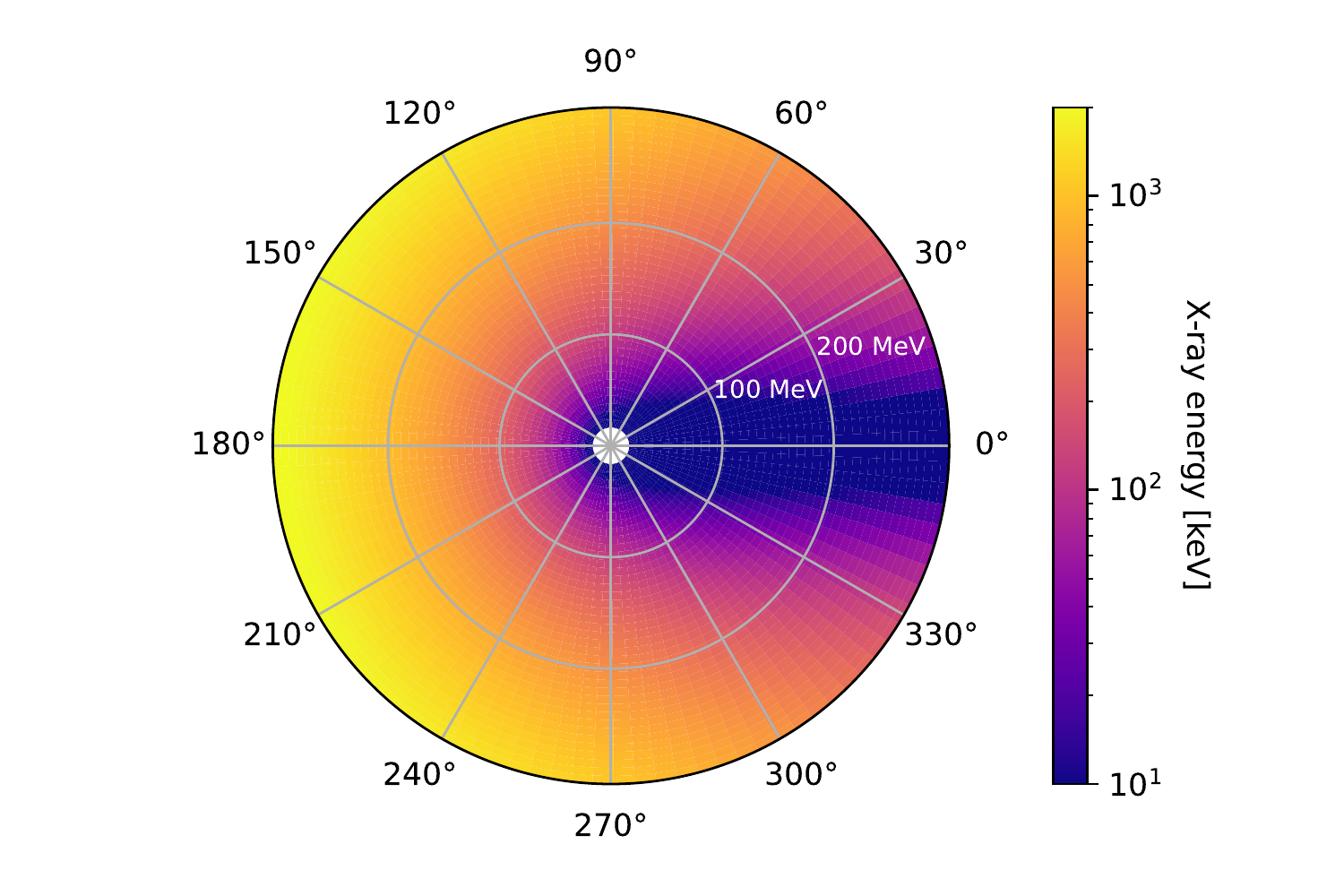}
\caption{Visualization of the scattered X-ray photon energy for different electron beam energies and scattering angles based on \cref{eq:PhotonEnergy}.}
\label{Compton_angles}
\end{figure}

Note that this expression has the same origin as the energy shift related to the observation angle $\theta$, i.e. the relativistic doppler shift. \Cref{Compton_angles} shows the radiated x-ray energy as a function of the collision angle and the electron energy. As discussed above, the x-ray photon energies produced by electrons at $100-200$ MeV are above a few hundreds of keV in the case of a head-on collision. However, upon closer inspection we can observe that the desired photon energies below 100 keV can be produced by changing the scattering geometry to a shallow angle of incidence $\varphi\lesssim\SI{30}{\degree}$. In this configuration the fundamental energy is down-shifted by a factor of $\sim 15$ or more with respect to head-on collision, meaning that scattering with electron beams of 100 to 250 MeV results in the emission of X-rays in the range of 15 to 100 keV. 

In the following, we are going to numerically investigate the potential of such a modified scattering geometry for narrowband X-ray generation. The paper is structured as followed, first we present results of simulations for Compton scattering for different scattering angle and energy combinations (\cref{Results}) that theoretically yield the same backscattered photon energy. These simulations are performed for electron beams of fixed divergence and emittance, respectively. We then present results for a fixed scattering angle ($\SI{30}{\degree}$) and tunable electron beam energies. Last, we discuss the properties of the radiated photon beams in the context of applications (\cref{Tomo}), with particular emphasis on the mitigation of beam hardening in computed tomography, and summarize our results (\cref{Conclusion}). 

\subsection{Results}\label{Results}

As shown in \Cref{Compton_angles}, electron beams with different energy $\gamma$ should emit the same X-ray energy $\hbar\omega$ if the scattering pulse arrives at an angle (in radian)
\begin{equation}\label{eq_angles}
    \varphi = 2\arcsin\left(\frac{1}{2\gamma}\sqrt{\frac{\hbar \omega}{\hbar\omega_0}}\right)\approx \frac{1}{\gamma}\sqrt{\frac{\hbar \omega}{\hbar\omega_0}}.
\end{equation}
Taking a target energy of $\SI{100}{\keV}$, which corresponds to an upshift of about 64500 compared to the laser energy $\hbar\omega_0=\SI{1.55}{\electronvolt}$, we find that the scattering angle is approximately given as $\varphi\approx \SI{75}{\degree}/E[\SI{100}{\MeV}]$. In the following we will thus analyze several different combinations of electron energy and scattering angles that should, in theory, yield to X-ray emission at the same energy.

\begin{figure*}
     \includegraphics[width=.94\linewidth]{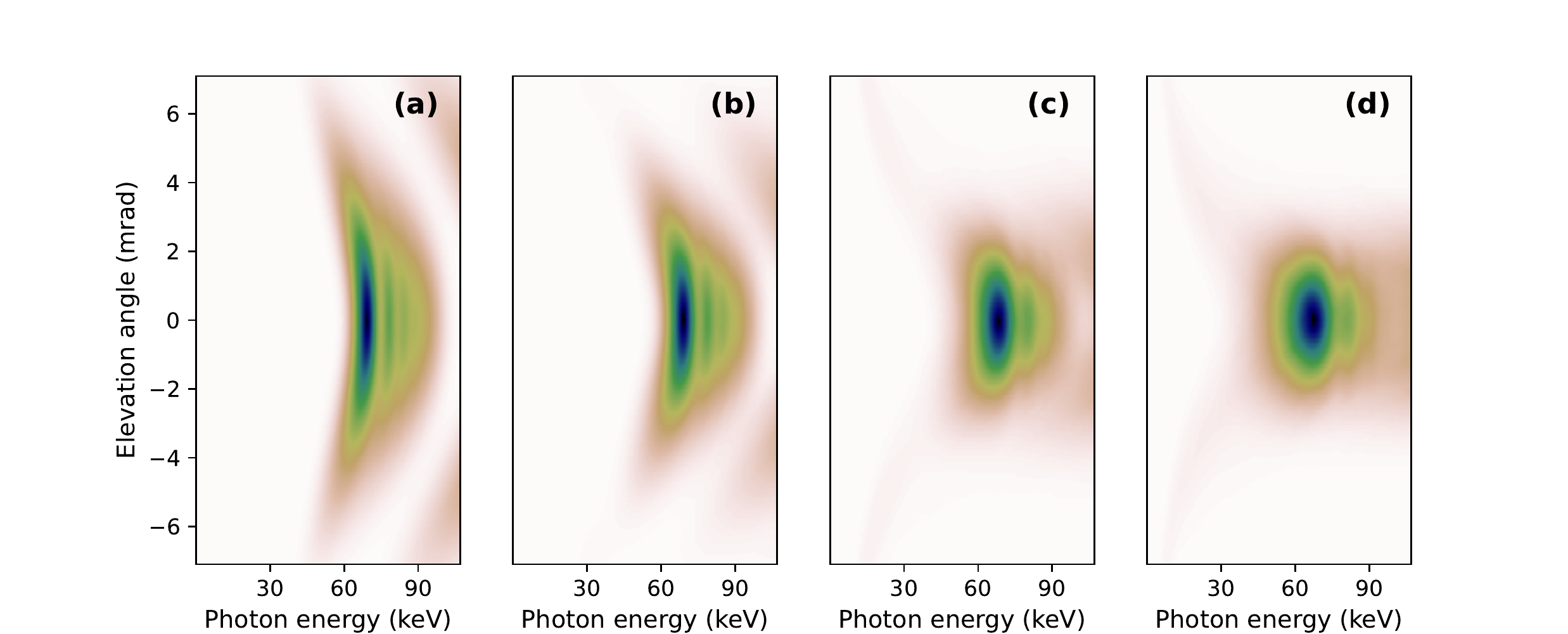}  \includegraphics[trim=0 -30 0 0, clip,width=0.045\linewidth]{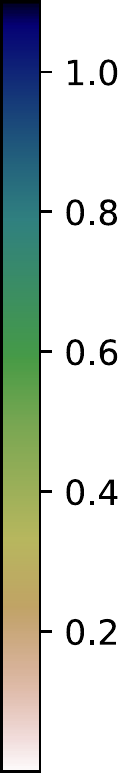}
     \includegraphics[width=.94\linewidth]{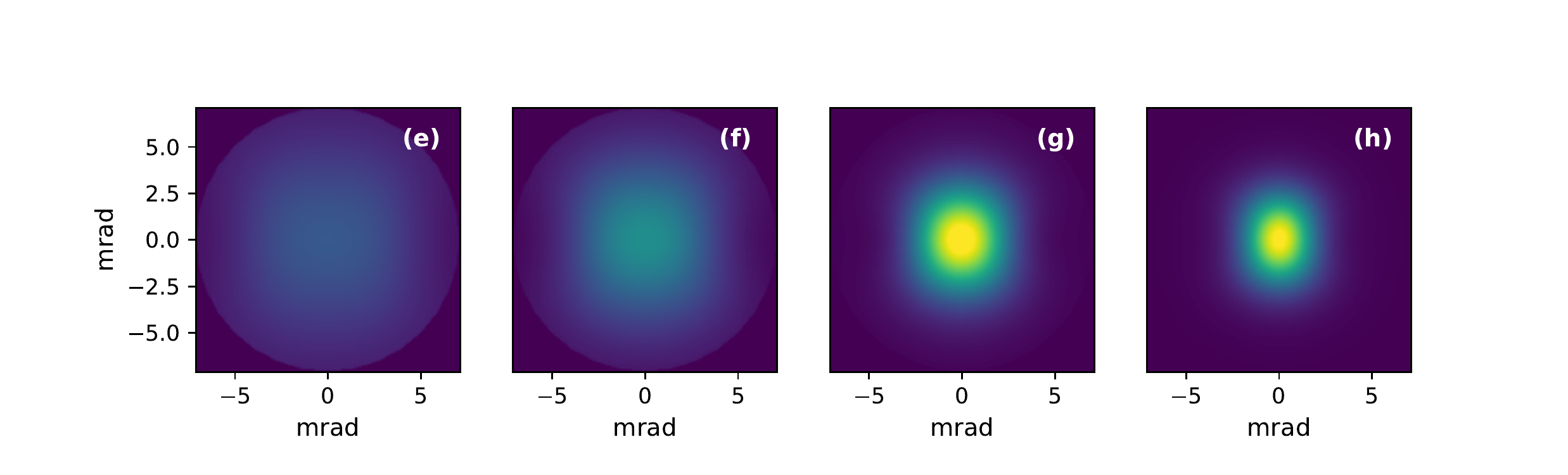} 
\caption{Angular-spectral (a-d) and 2D angular (e-h) distributions of emitted energy and angular Compton scattering at 70 keV peak photon energy produced with electron energies 65~MeV, 92~MeV, 170~MeV, 250~MeV and scattering angles 180$^\circ$ 90$^\circ$ 45$^\circ$ 30$^\circ$ respectively. Colormaps (a-d) are normalised individually, colormaps (e-h) have a common normalization.}
\label{Compton_angles_2}
\end{figure*}

We have simulated four different scattering cases by reconstructing electron trajectories in a laser pulse with the Gaussian temporal and spatial profiles using an implicit 5th order Radau IIA family Runge-Kutta method and calculating the emitted radiation field using the \textsc{SynchRad} software \cite{Andriyash}. The electron bunch was initiated with 8000 test particles with 1 percent energy spread and 1 mrad divergence, and the laser had the amplitude $a_0=1.0$ (weakly non-linear), 30 fs duration and $\SI{20}{\micro\meter}$ FWHM spot size at the interaction point. The electron beams have energies of 65 MeV, 92 MeV, 170 MeV and 250 MeV, with respective scattering angles of 180$^\circ$, 90$^\circ$, 45$^\circ$ and 30$^\circ$. In all cases laser polarisation was oriented perpendicular to the plane in which electron and laser beams cross, to assure a more efficient linear polarisation of the scattered radiation.

The resulting emission characteristics are shown in \cref{Compton_angles_2}. The simulations confirm that all four simulations scenarios yield the same X-ray energy. One result, however, is that the X-ray energy is centered at about $\SI{70}{\keV}$ instead of $\SI{100}{\keV}$. This is readily explained by the fact that our derivations were based on the assumption of $a_0\ll1$, while the simulations use the more realistic case of an intense scattering pulse with $a_0=1.0$. This results in a downshift to about $1/(1+a_0^2/2)=2/3$ of the target energy (cf. \cref{eq1}). The latter is only an approximation of the downshift because the electrons do not interact all the time with the peak potential $a_0=1$, but also scatter on the slopes where $a_0<1$. As such, looking at the on-axis emission, we can observe radiation from 66 keV (corresponding to $a_0=1$) up to 100 keV ($a_0\ll1$). Most of the emission occurs at an 
 effective scattering potential $a_{0, eff}$, which depends on the pulse shape and the shift can be accounted for by adding this dependence\footnote{The corresponding, modified version of \cref{eq_angles} is $\varphi = 2\cdot \arcsin\left(\frac{1}{2\gamma}\sqrt{\frac{\hbar \omega}{\hbar\omega_0}\cdot{(1+a_{0,eff}^2/2)}}\right)$.}. 

\begin{figure*}
     \includegraphics[width=.94\linewidth]{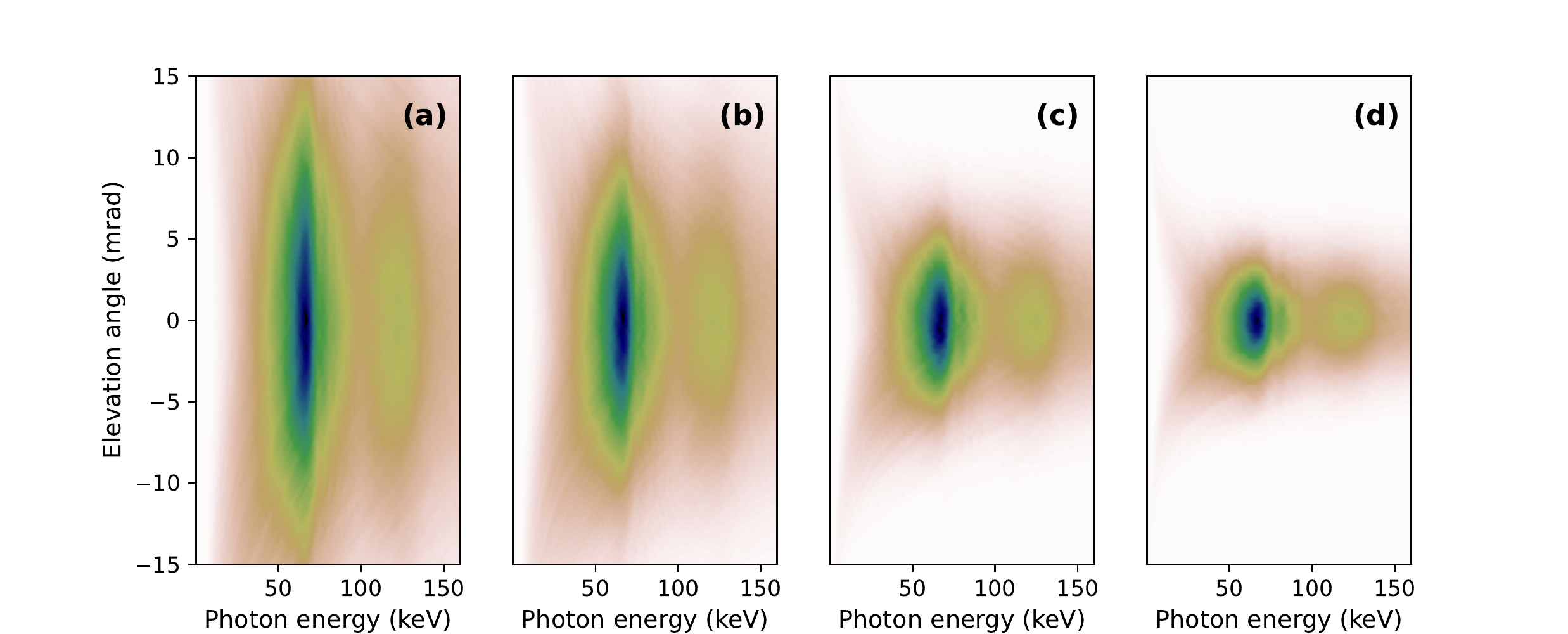}  \includegraphics[trim=0 -30 0 0, clip,width=0.045\linewidth]{Images/cbar_sample_vert.pdf}
\caption{Angular spectral distributions of emitted energy of the Compton scattering at 70 keV peak photon energy produced with electron energies of constant emittance and otherwise same parameters as in \cref{Compton_angles_2} (65~MeV, 92~MeV, 170~MeV, 250~MeV and scattering angles 180$^\circ$, 90$^\circ$, 45$^\circ$ and 30$^\circ$, respectively).}
\label{3_Constant_emittance}
\end{figure*}

Upon closer examination we observe that the angular-spectral distributions resemble the well-known horse-shoe shape of an undulator for low-energy cases, dictated by the $(1+\theta^2\gamma^2)^{-1}$ term in \cref{eq1}. However, at higher electron energies the distribution becomes more blurred. This is because X-ray the emission cones of electrons with different divergence start overlapping. Starting again from Eq.\ref{eq1}, we can estimate the local bandwidth as the energy difference between photons emitted on axis and those coming from an electron crossing at a divergence angle $\sigma_\theta$. Replacing the observation angle term $\theta\gamma$ from \cref{eq1} with this, the local energy bandwidth is approximately given by 
\begin{equation}\label{bandwidth}
\hbar\Delta\omega \approx 4\gamma^2\sin^2\frac{\varphi}{2} \frac{(\sigma_\theta\gamma)^2}{1+(\sigma_\theta \gamma)^2}.
\end{equation}
Thus, a 1-mrad divergence of electrons at the point of scattering leads to a relative energy bandwidth of 3.5\% in the case of a 100 MeV beam, but 20\% for a 250 MeV beam. 

A further contributing factor to the bandwidth is the beam energy spread $\Delta \gamma$. We can estimate this effect by plugging $\gamma\pm \Delta \gamma$ into \cref{eq1}, which yields to a relative difference of the backscattered energy of $2\Delta \gamma / \gamma$. Last, one should mention that even for a perfectly collimated, monoenergetic electron beam scattering with a laser with rectangular intensity profile, the bandwidth is fundamentally limited by the number of oscillations $N_{osc}$ to $\hbar\omega/N_{osc}$. While this is not a limiting factor for our conditions, it is worth noting that the oscillation number is actually dependent on both the laser pulse duration and waist for $\varphi\neq \SI{180}{\degree}$. This is discussed in more detail in the appendix.

\begin{figure}[t]
\includegraphics[width=0.98\linewidth]{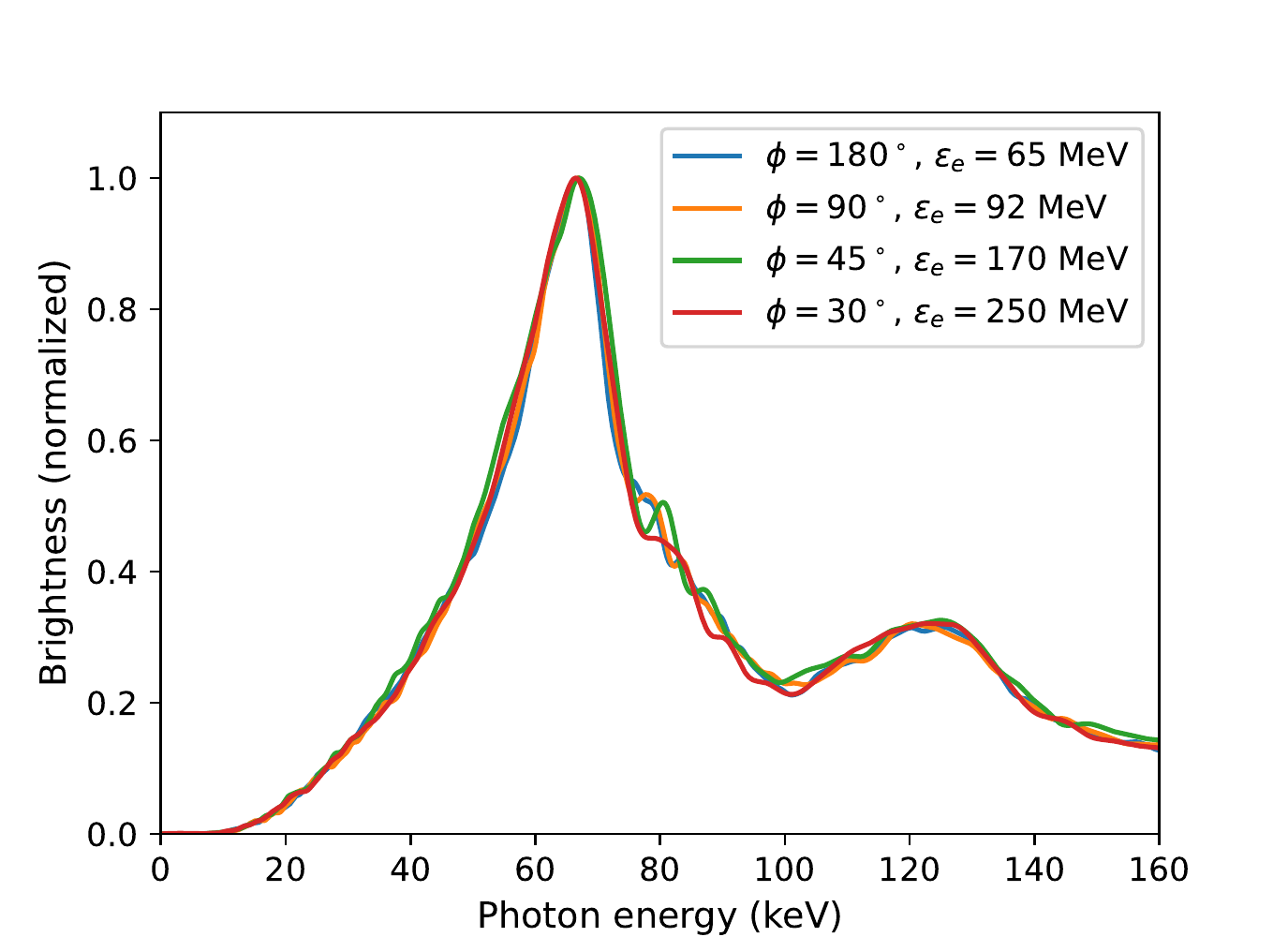}
\caption{On-axis lineout of the spectra shown in \cref{3_Constant_emittance}.}
\label{4_On_axis}
\end{figure}
\begin{figure}[!b]
\includegraphics[width=0.93\linewidth]{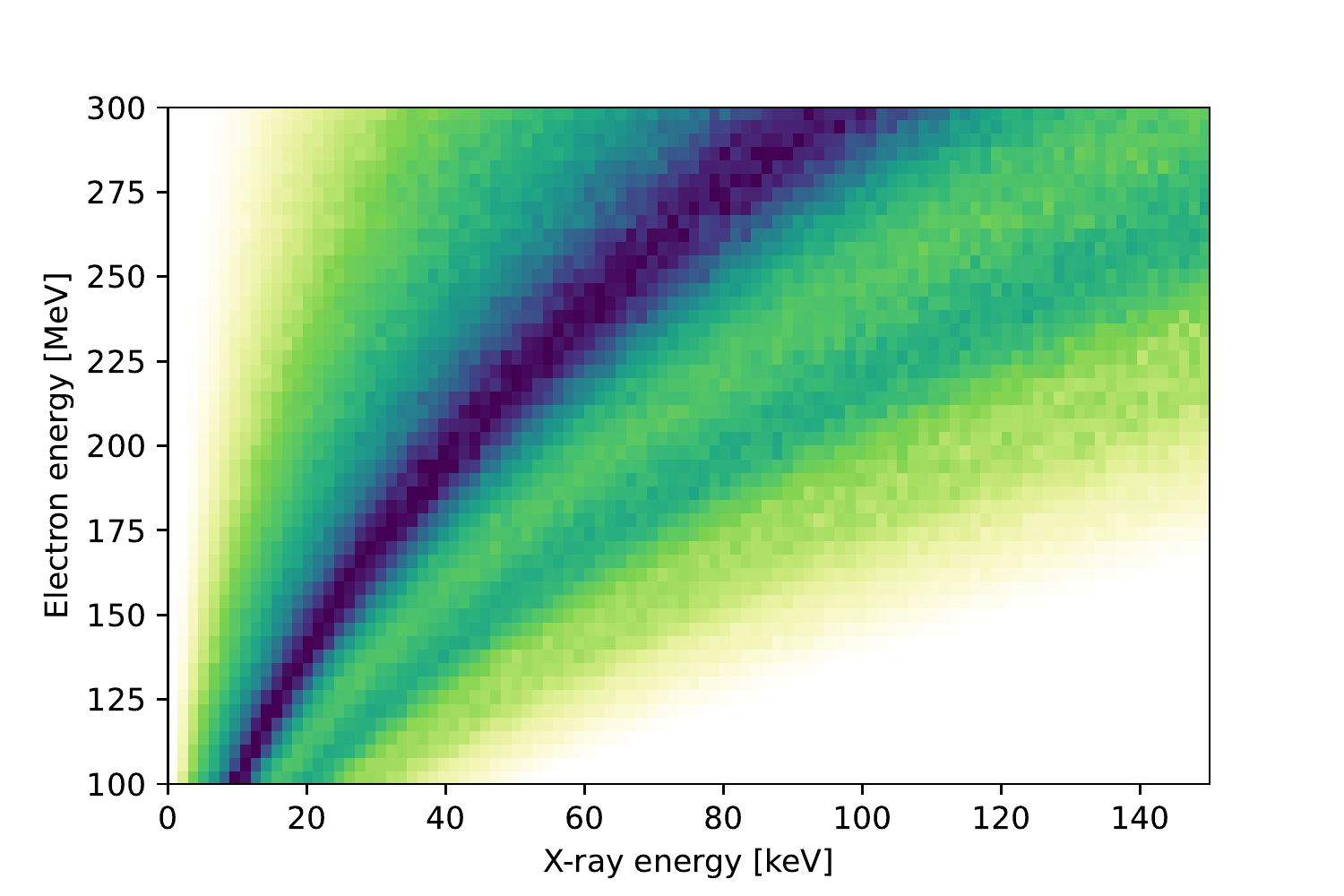}
\caption{On-axis spectral distribution of emitted energy of the Compton scattering energies from 100 to 300 MeV. Calculated based on \cref{eq1} including the first four harmonics. The emitted power at each electron energy is normalized for better visibility of the spectral form.}
\label{5_Compton_energies}
\end{figure}

These four effects, varying $a_0$, energy spread, electron divergence and number of oscillations, contribute to the bandwidth observed in the simulations, whose on-axis spectral bandwidth grows accordingly from 10\% to 25\% from (a) to (d) of \cref{5_Compton_energies}. The $a_0$ contribution is difficult to estimate, but from our analysis we find that the contribution of beam divergence largely outweighs the energy spread, making this the more important electron beam parameter to optimize for narrowband emission. In fact, laser wakefield accelerators would typically produce electron beams with a significantly larger divergence than 1 mrad at 65 MeV. This is because the acceleration process conserves the normalized emittance of an electron beam $\epsilon_0 = \gamma\sigma_{\theta}\sigma_x$, which is typically of the order of $\epsilon_0 =\SI{1}{\mm}.\si{\milli\radian}$. 

For a more realistic comparison, we have thus repeated the simulations for beams with a divergence adjusted to this emittance instead of the previously constant divergence. The results are summarized in \cref{3_Constant_emittance} and \cref{4_On_axis}. The corresponding divergence changes from  $\SI{8}{\milli\radian}$ for $\SI{65}{\MeV}$ and down to $\SI{2}{\milli\radian}$ for $\SI{250}{\MeV}$. As we maintain the same beam size $\sigma_x$, the factor $\gamma\sigma_{\theta}$ is constant in this case and from \cref{bandwidth} we expect the same energy bandwidth for each electron energy. Indeed, as shown in \cref{4_On_axis} the shapes of the on-axis spectra are identical in these cases, while the angular spectral distributions (cf. \cref{3_Constant_emittance}) appear stretched or squeezed proportional to $1/\gamma$.

\begin{figure*}[hbt]
\includegraphics[width=1\linewidth]{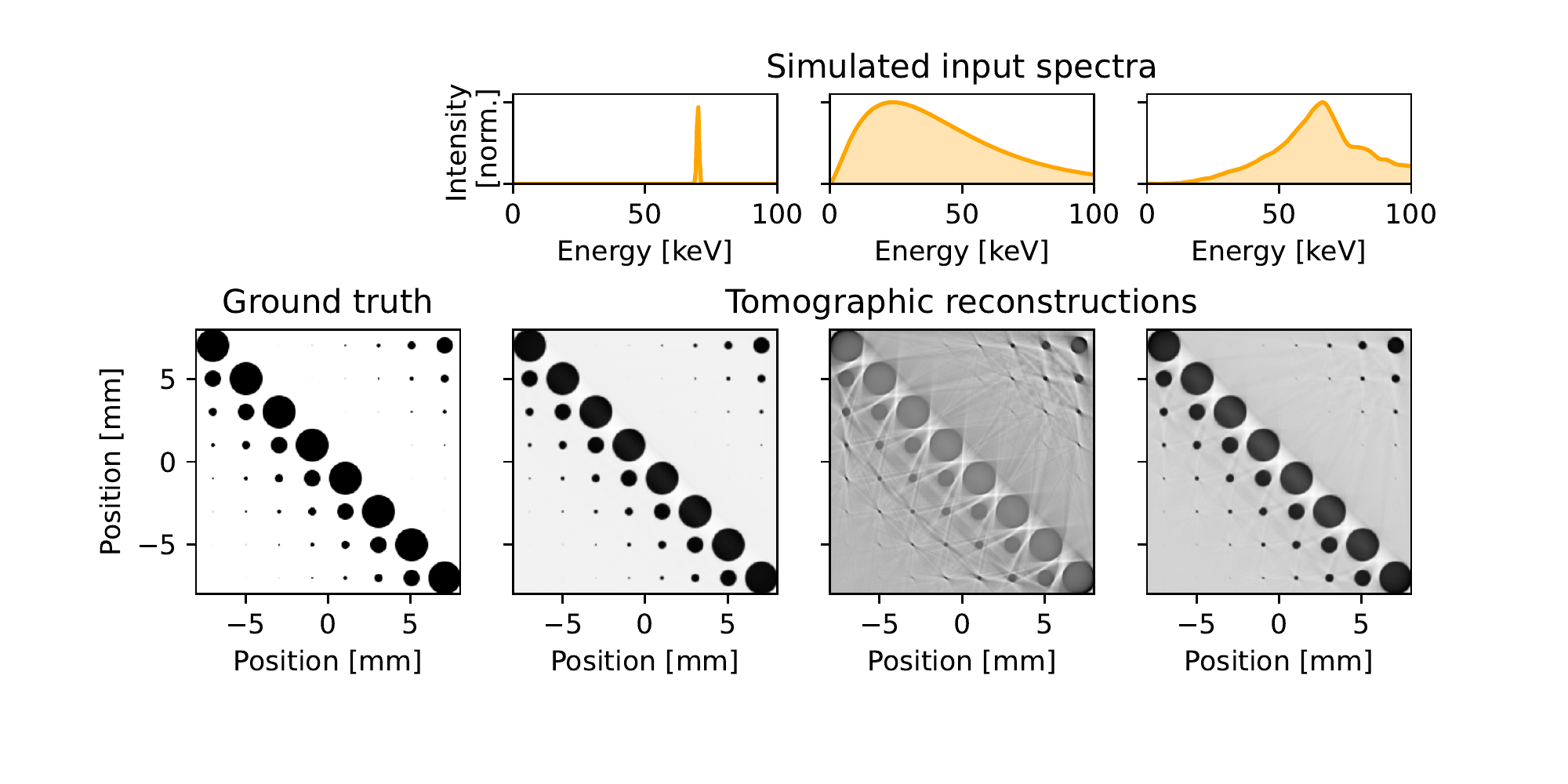}
\caption{Simulated tomography to estimate the extend of beam hardening artifacts given different X-ray spectra. The ground truth, a slice through an aluminum rod array, is shown on the very left. What follows from left to right are input spectra and reconstructions for a monoenergetic 70 keV source, a betatron spectrum with 40 keV critical energy and the simulated on-axis Compton spectrum from \cref{4_On_axis}.}
\label{Tomography}
\end{figure*}

In all cases simulated so far, the full emitted energy is nearly conserved and estimates as $5$~nano-Joules per a pC charge. The change in the X-rays angular divergence in this case affects the angular energy density (integrated brightness) as presented in \cref{Compton_angles_2}(e-h). This is expected from the scaling of the radiation cone with $1/\gamma$ and thus, scattering with higher electron beam energies leads to more collimated X-ray emission.

From the previous sections it thus appears that scattering at a shallow angle such as $\varphi=\SI{30}{\degree}$ leads to more collimated beams than head-on collision, while otherwise maintaining the same beam parameters. This makes it promising to combine this configuration with an energy-tunable laser wakefield accelerator. In the following we thus present results for scattering with electron beams varying in energy from $100-300$ MeV, which is a tuning range that can be covered via various proven technologies (shock-front injection \cite{Buck.2013}, tunable gas-cells \cite{Corde.20133vm}, etc.). Given the good agreement between simulations and analytical scalings for on-axis emission (see \cref{comp}), we use an estimation based on \cref{eq1} here, allowing us to calculate a continuum for emissions in the energy range. 
The calculation results presented in \cref{5_Compton_energies} demonstrate that using $100-300$ MeV electrons such configuration can provide the narrow-bandwidth radiation varying in the range from 10~keV to 100~keV, which is the relevant regime for most imaging applications.

\subsection{Potential for applications in X-ray tomography}\label{Tomo}

The quasi-monochromatic nature of the X-ray beams generated via Compton scattering makes them interesting for medical imaging and non-destructive testing. As we have seen in the previous part, the X-ray bandwidth is dominated by two factors, the scattering pulse shape and the electron beam divergence according to \cref{bandwidth}. For a given emittance, the on-axis spectrum of the backscattered X-rays is unaffected by a change of the scattering angle according to \cref{eq_angles}. Meanwhile, the angular distribution scales with $1/\gamma$, leading to more collimated X-rays for scattering at shallow angles and high electron energies. Regarding a potential application in X-ray imaging, this has both positive and negative consequences. A more collimated beam is easier to transport and produces near-parallel projections that are easier to handle than fan beam projections \cite{Kalender.2006, Withers.2021}. However, to illuminate larger objects the distance to the source needs to be increased, which is not desirable for compact setups. More precisely, given an object size $D$ and an emission cone of $1/\gamma$, the minimum distance for imaging is $d_{min}\simeq \gamma \cdot D$.

An interesting feature of the Compton signal from electron beams with non-zero divergence is the aforementioned mixing between emissions at different angles. While this causes an increase in bandwidth, cf. \cref{bandwidth}, it also strongly reduces the angular dependence of the spectrum. This is beneficial for imaging, as it assures that the absorption of samples is the same over the entire field of view. 

One particular advantage of narrowband sources is that they do not produce beam hardening artifacts in tomographic reconstruction \cite{Boas.2012}. The latter are artifacts that originate from unequal absorption of different X-ray energies in a sample. We can briefly estimate to which extend the beams simulated in the previous sections would produce such artifacts and compare this with synthetic tomograms based on monochromatic sources and broadband spectra from laser-driven betatron radiation as used in \citet{Cole.2015n7b} or \citet{Doepp.2018wil}. To do so, we generate synthetic tomograms based on the 'Doga’s Circles' phantom from the TomoBank database \cite{Carlo.2018}. Consisting of rods with different diameter that align for certain projection angles, this phantom is particularly well suited to study beam hardening. We have adjusted the phantom such that the largest rod diameter corresponds to $\SI{2}{\mm}$ of aluminum. The sample area encompasses $400\times 400$ pixels and we calculate 400 parallel projections over a range from 0 to $\SI{180}{\degree}$. We then calculate the X-ray transmission based on the tabulated transmittance of aluminum \cite{Berger.1999j0f} and the spectrum of a simulated input source, resulting in a transmission sinogram. For simplicity, we assume a detector with perfect quantum efficiency in this step. From this sinogram we calculate the spatially-resolved transmittance via filtered backprojection.

The results are shown in \cref{Tomography}. As expected, the reconstruction using a monochromatic spectrum of 70 keV X-rays does not show any artifacts. However, the reconstruction using a betatron input spectrum with a critical energy of 40 keV is heavily distorted by beam hardening artifacts. These artifacts are still present using a realistic Compton input spectrum due to the low-energy tail in the spectrum, but the artifacts are heavily reduced. Applying weak spectral filtering of the X-rays or using an electron beam with even smaller emittance could further suppress these artifacts, while essentially keeping the X-ray flux the same. Alternatively, one could perform dual-energy tomography by tuning the electron beam energy, which will also drastically reduce beam-hardening artifacts \cite{Coleman.1985}.

\subsection{Conclusion and outlook}\label{Conclusion}

To conclude, optical undulators offer the possibility to tune the backscattered energy not only by means of the undulator period and the electron energy, but also by using different collision angles $\varphi$. When operating at small $\varphi$, the beam collimation is increased with respect to equivalent counter-propagating Compton-sources. While this makes the source's spectrum in principle more sensitive to the electron beam divergence, we find that this effect compensated due to normalized emittance conservation.
The properties of the emitted radiation are of interest for applications such as non-destructive testing and medical imaging. We have identified a shallow scattering angle of $\SI{30}{\degree}$ is particularly interesting for the LWFA-based all-optical configurations and plan to test this geometry in future experiments.

\section{APPENDIX}

\subsubsection{Influence of the collision angle on the interaction time}

\begin{figure}[hbt]
\includegraphics[trim=100 110 100 110, clip,width=1\linewidth]{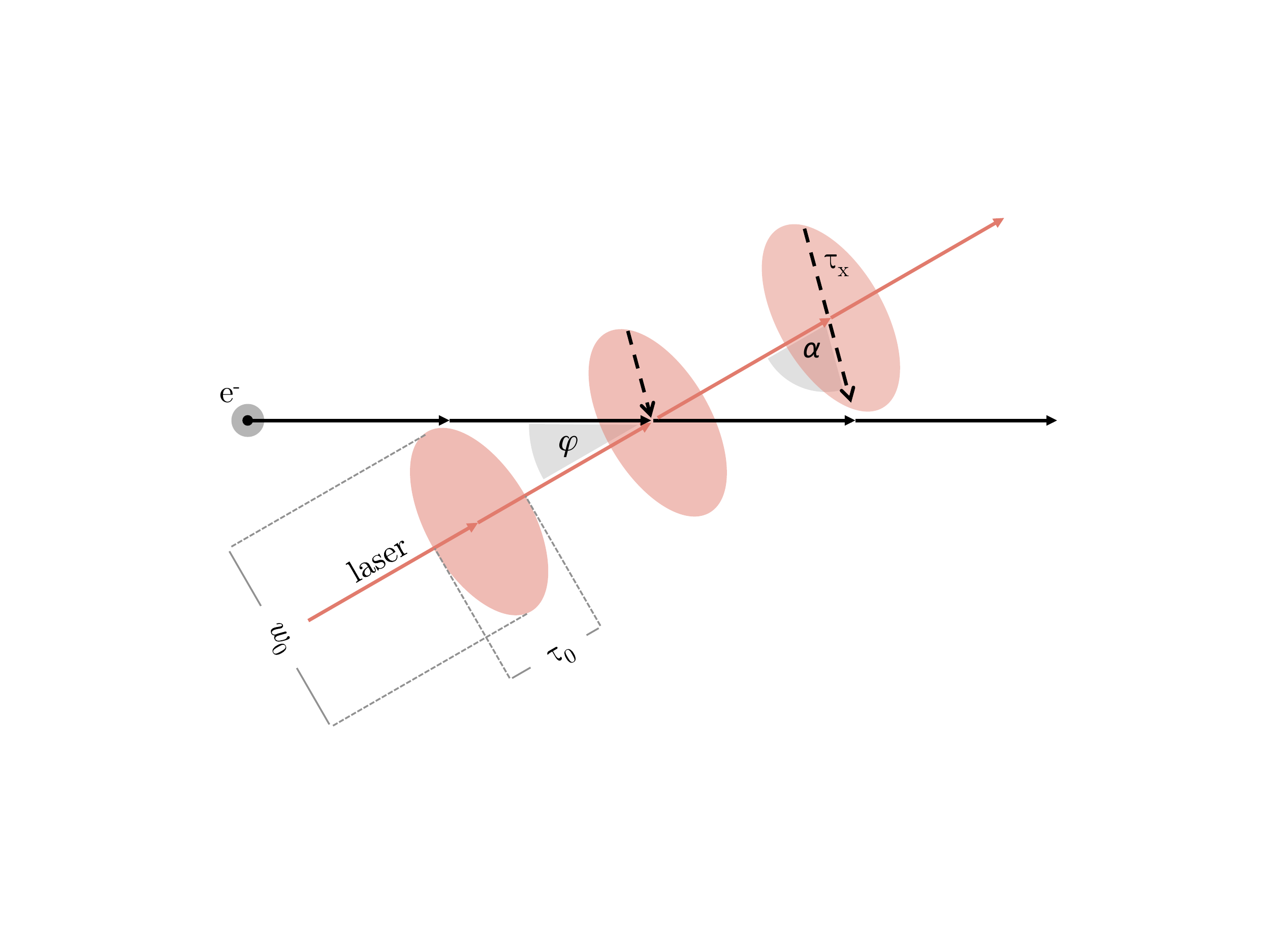}
\caption{Sketch of the scattering geometry.}
\label{Compton_geometry}
\end{figure}

When using different collision angles $\varphi$, it is also important to estimate the consequences for the interaction time. For this we consider a simple geometric model, shown in \cref{Compton_geometry}. We approximate the electron beam as point-like, while the laser beam is described as an ellipse with the diameters $a=\tau_0$ and $b=w_0/c_0$. The electron beam propagates at a velocity close to the speed of light $c_0$, so the laser beam ellipse is cut at an angle $\alpha = (\SI{180}{\degree} - \varphi)/2$. From the polar form of the ellipse equation we then get  
$$\tau_x(\varphi) = {ab}\left[(b\cos\alpha)^2+(a\sin\alpha)^2\right]^{-1/2}.$$
For $\varphi=\SI{180}{\degree}$ this yields as expected $\tau_x=\tau_0$, meaning that the interaction time is given by the laser pulse length. In co-propagating geometries ($\varphi\rightarrow \SI{0}{\degree}$) the laser waist is the defining parameter. Also, for a given waist $w_0$ and $\tau\rightarrow\infty$, the interaction time is $\tan\alpha \times w_0/c_0$. At $\varphi=\SI{30}{\degree}$, for example, this value is about $3.7w_0/c_0$, so at a typical waist of 20 microns the interaction time cannot exceed $\sim 250$ fs. It is important to note that a quasi-co-propagating scattering geometry may lead to a longer interaction time, but never increases the number of oscillations. This is an important issue because the amount of photons emitted depends on the number of laser cycles $n$ at the wavelength $\lambda_0/\sin^2(\varphi/2)$ the electrons perceive (e.g. $n_{max}\sim 10$ for the afore-mentioned case). To limit the efficiency loss because of the reduced number of oscillations, one may spatially tilt the laser wavefront, as proposed in schemes such as the travelling-wave undulator \cite{Debus.2010}. Depending on the tilt direction, the laser will then cross the electron beam at $\alpha=\SI{0}{\degree}$ or $\alpha=\SI{90}{\degree}$ and the interaction time is $\tau_0$ or $w_0/c_0$, respectively. However, this implementation is challenging regarding its physical implementation. 

\subsubsection{Comparison between simulations and analytical results}\label{comp}

\begin{figure}[hbt]
\includegraphics[width=1\linewidth]{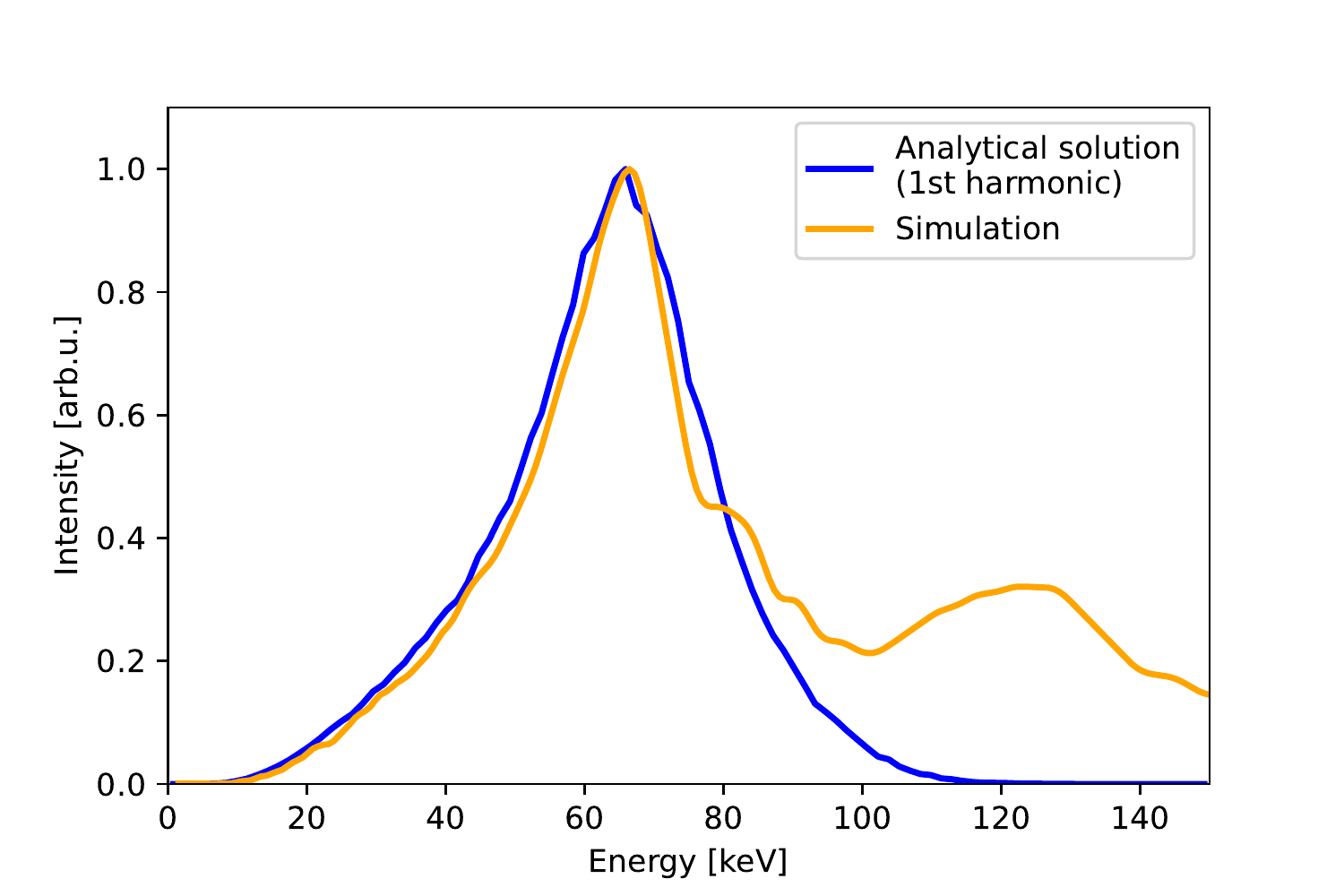}
\caption{Comparison between simulations and analytical results based on \cref{eq1}.}
\label{Comparison}
\end{figure}

In the manuscript we extensively use \cref{eq1} and simplifications of it to predict properties of the X-ray beam. To justify this, we compare the results shown in \cref{4_On_axis} with a numerical solution of \cref{eq1}. For this numerical solution we solve the equation for different time steps along a pulse with Gaussian profile ($a(t)=a_0\cdot\exp(-(t/\tau)^2)$) and for 5000 electrons whose interaction angle (divergence) and energy are varied as in \cref{4_On_axis}. Furthermore, we assume $N_{osc}=10$ and sample the final energy from a normal distribution with width $\hbar\gamma/N_{osc}$. Last, to account for the difference in emitted power at different observation angles, we scale the emitted power with $(\gamma^{-2}+\theta^{2})^{-{3/2}}$. Note that this factor was ignored in the analytical analysis. As shown in \cref{Comparison}, the predicted spectral shape of the fundamental emission agrees well with the simulation.

\end{document}